\documentclass[11pt,a4paper]{article}

\usepackage[margin=27mm]{geometry}
\usepackage{fontspec}
\usepackage{xeCJK}
\xeCJKsetup{CJKspace=true}   
\setCJKmonofont{UnBatang.ttf}
\newcommand{\rgbox}[1]{\raisebox{-0.12em}{#1}}
\newcommand{\rgYubin}{\rgbox{\includegraphics[height=1em]{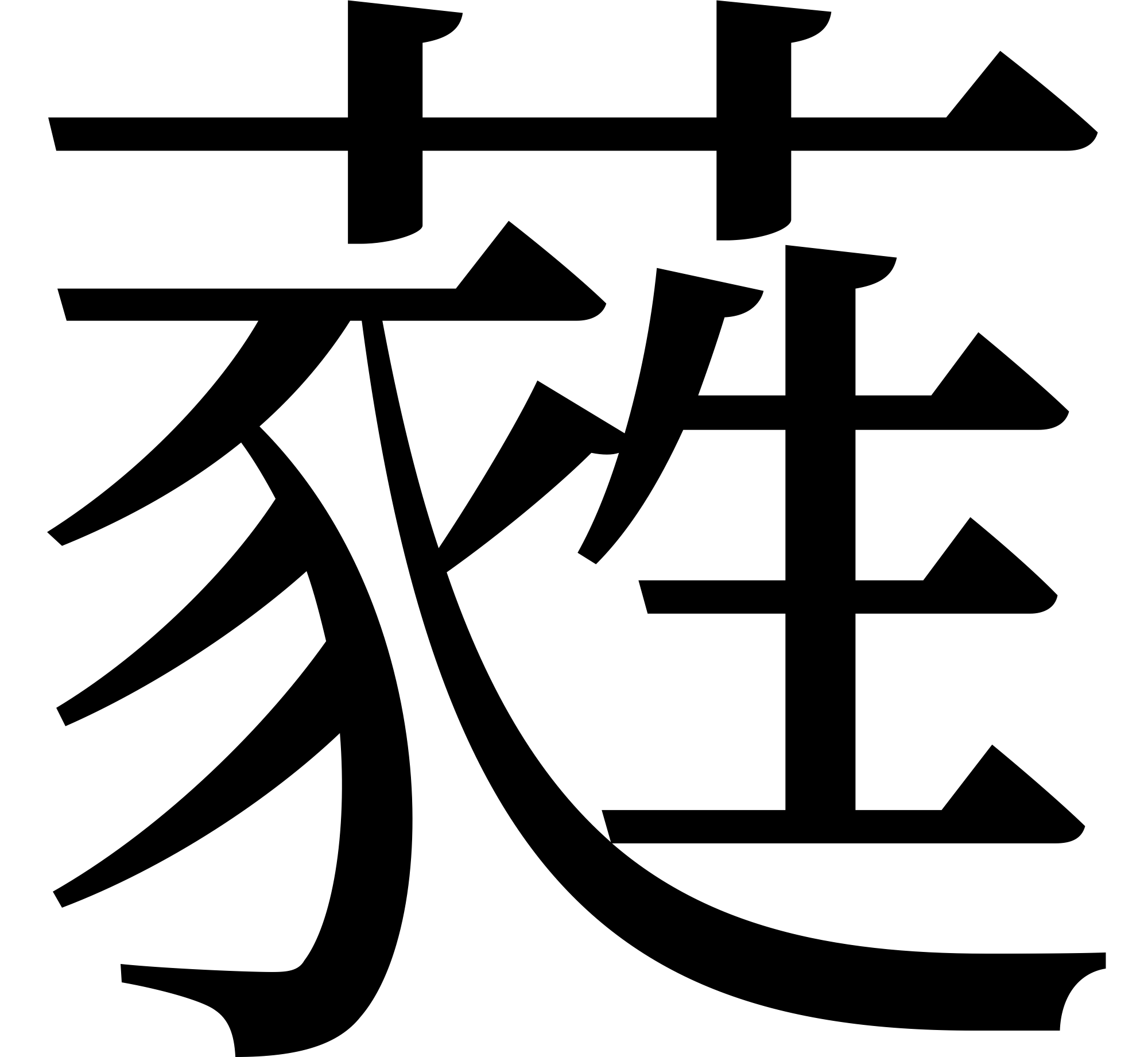}}}      
\newcommand{\rgSamsu}{\rgbox{\includegraphics[height=1em]{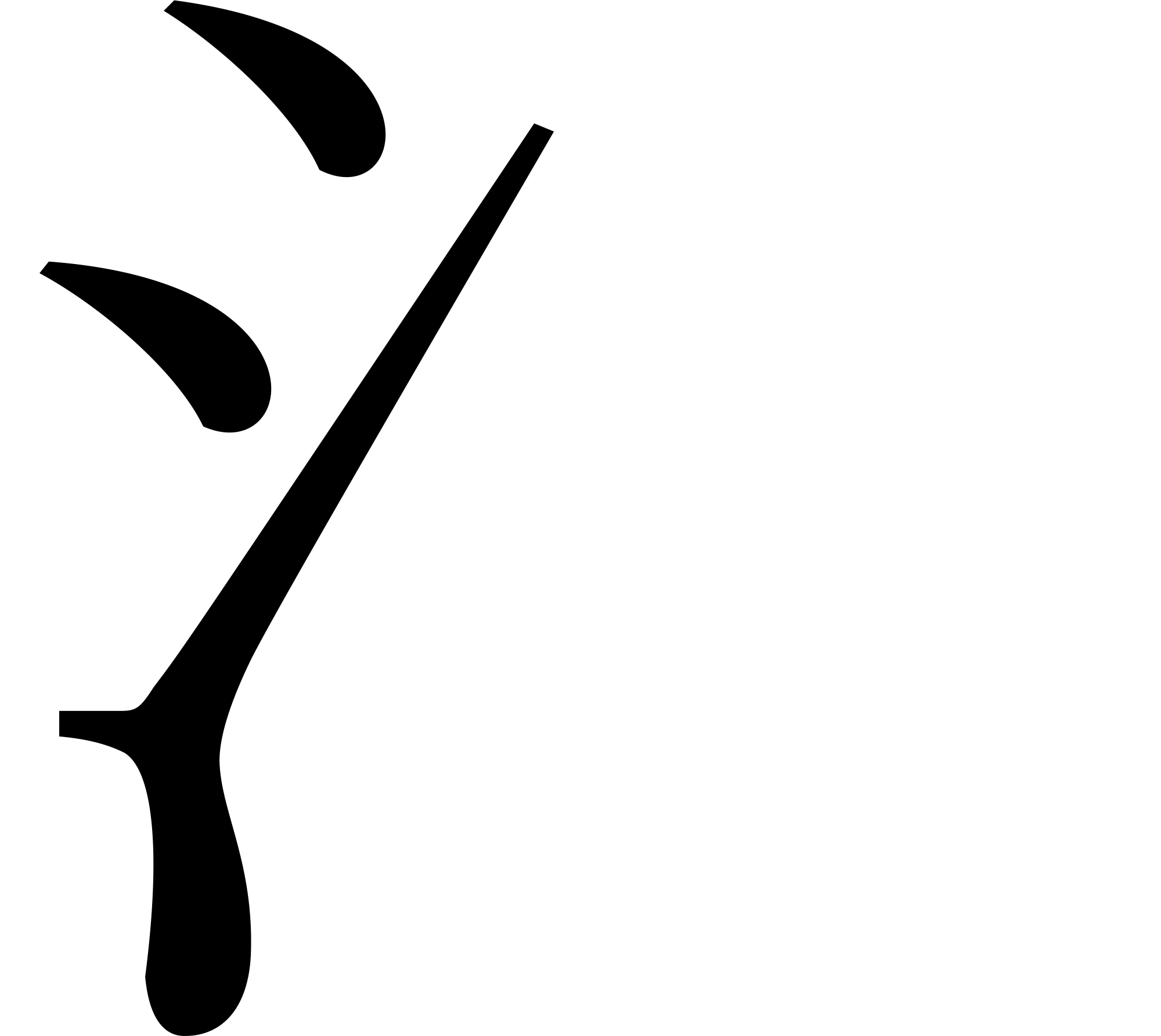}}}      
\newcommand{\rgHwangUpTwo}{\rgbox{\includegraphics[height=1em]{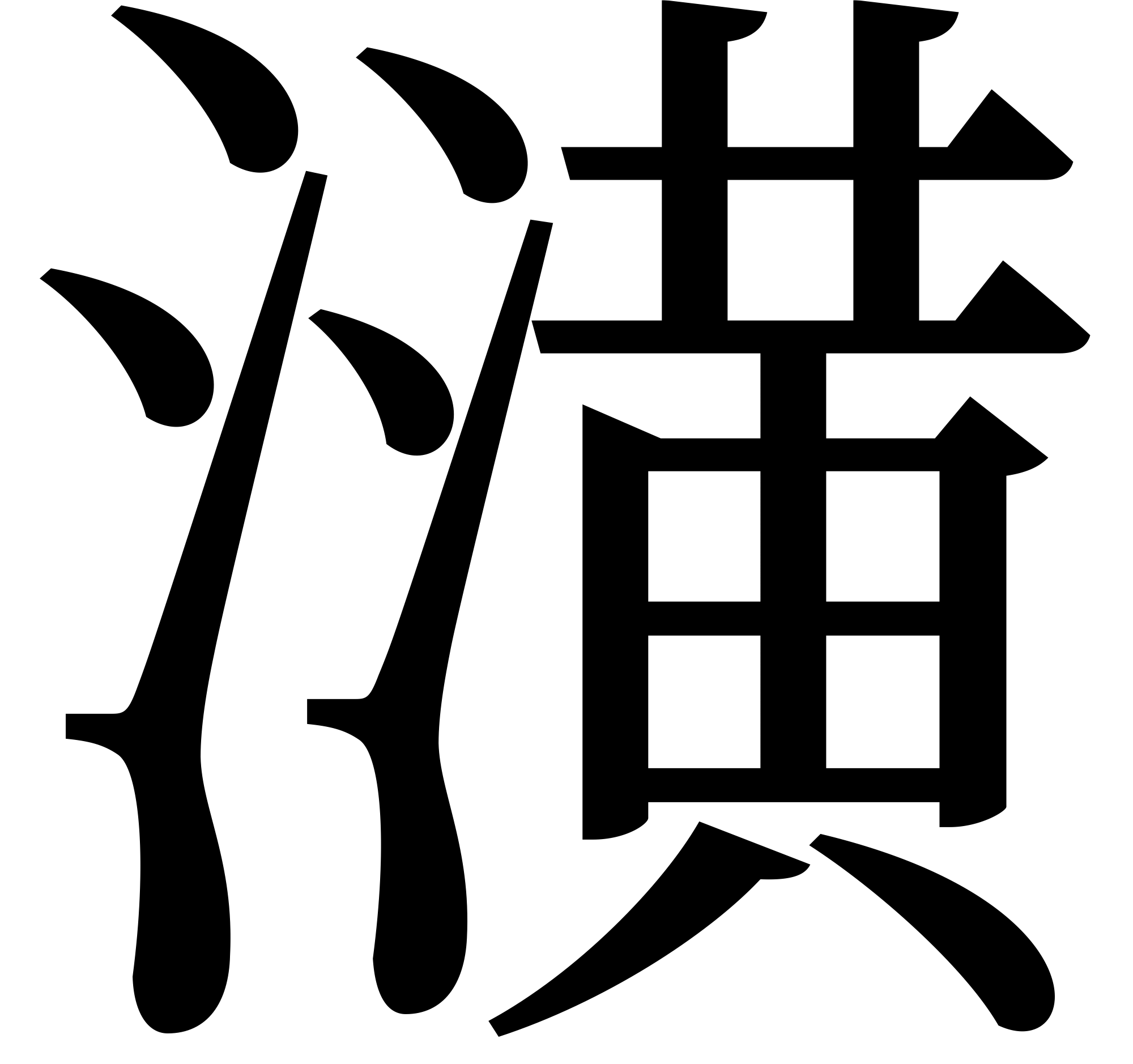}}} 
\newcommand{\rgInbyeon}{\rgbox{\includegraphics[height=1em]{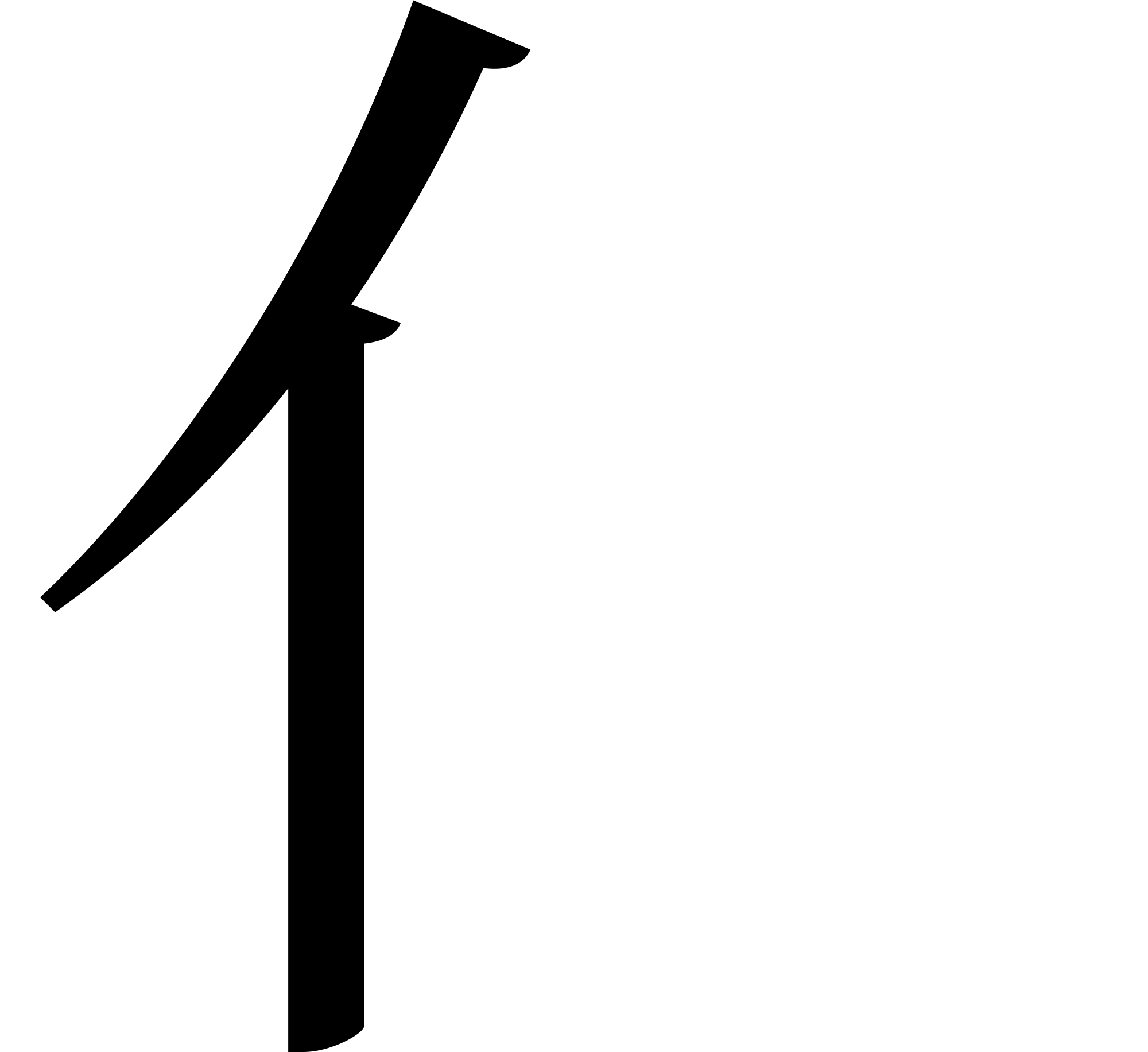}}}    
\newcommand{\rgHwangDownOne}{\rgbox{\includegraphics[height=1em]{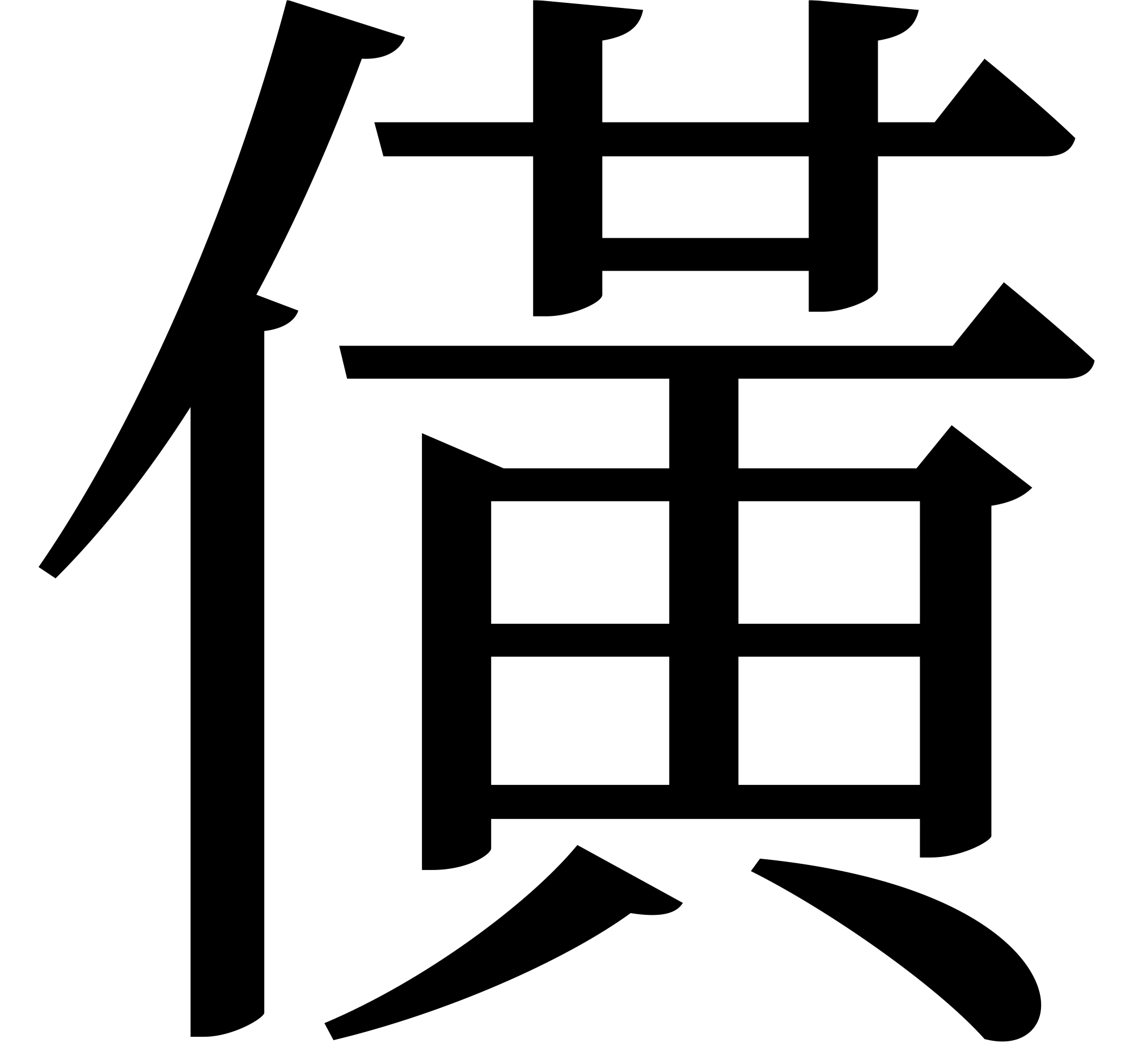}}} 
\newcommand{\rgDuin}{\rgbox{\includegraphics[height=1em]{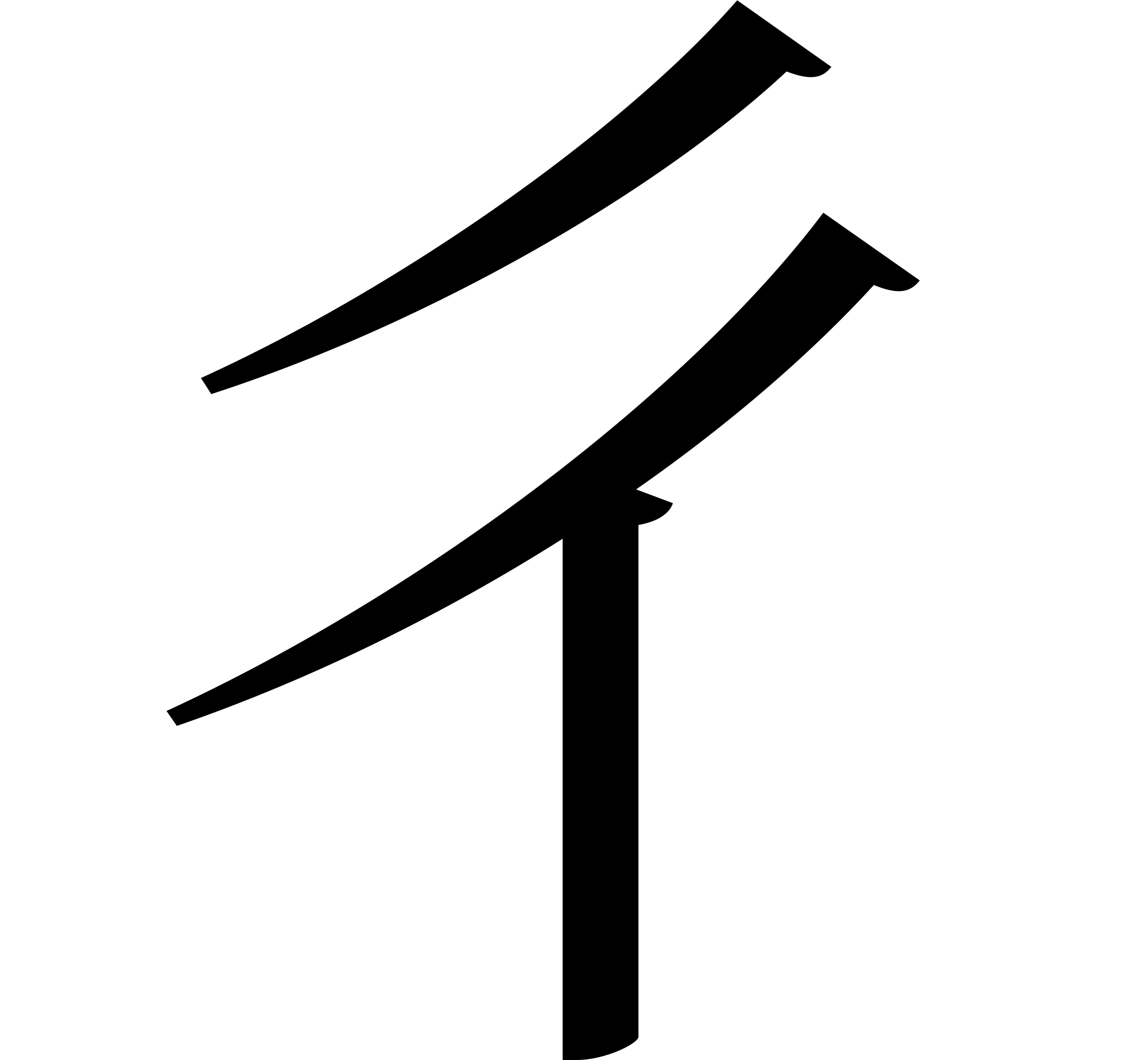}}}       
\newcommand{\rgHwangDownTwo}{\rgbox{\includegraphics[height=1em]{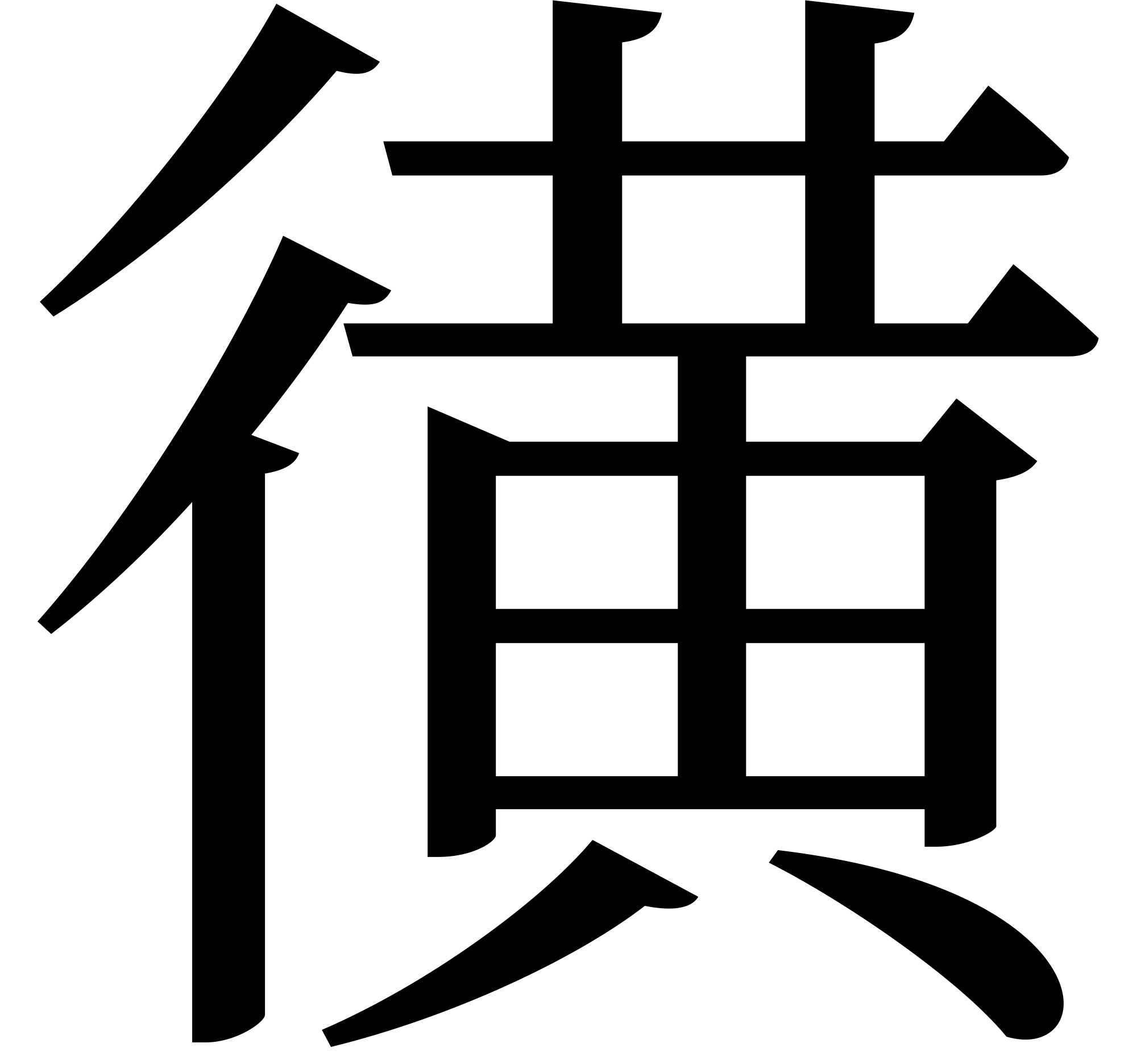}}} 

\usepackage{amsmath}
\usepackage{graphicx}
\usepackage{booktabs}
\usepackage{caption}
\usepackage{xcolor}
\usepackage[hidelinks]{hyperref}
\usepackage{enumitem}
\setlist{itemsep=2pt,topsep=4pt}
\newcommand{\kr}[1]{#1}                 
\newcommand{\term}[1]{\emph{#1}}        

\title{\textbf{Umulsai: A Plain-Text Format and In-Browser Engraving
System for \term{Jeongganbo}, Korean Mensural Notation}}
\author{Danbinaerin Han\thanks{Music and Audio Computing Lab (MAC Lab),
Graduate School of Culture Technology, Korea Advanced Institute of Science
and Technology (KAIST). \texttt{naerin71@kaist.ac.kr}}}
\date{Technical Report, v2026-07\\[2pt]
\normalsize System available at \url{https://www.umulsai.com}}

\begin{document}
\maketitle

\begin{abstract}
\term{Jeongganbo} (\kr{정간보}, 井間譜) is a grid-based mensural notation
created in fifteenth-century Korea and still in standard use for Korean
traditional music. Its structure --- a vertical column of square beat-cells
read right-to-left, in which the \emph{position and share} of a syllable
inside a cell encode its duration --- is incompatible with engraving systems
built for
Common Western Music Notation, and no practical digital tool for producing it
has been in sustained distribution. This report describes the technical design
of \term{Umulsai} (\kr{우물사이}), a notation editor for \term{jeongganbo}
that runs entirely in a web browser as a static, serverless application. Its
core is (i)~a human-writable plain-text format whose spatial arrangement
corresponds one-to-one with the notational grid --- line~= \term{gak} column,
delimiter~= \term{jeonggan} cell, whitespace~= beat subdivision, adjacency~=
intra-row placement --- so that duration is expressed by position and share in the text
exactly as it is on the page; (ii)~an engraving engine that formalizes the
typesetting conventions of published scores (right-to-left column flow,
\term{daegang} section lines, octave-variant Sinographs, ornament attachment
and micro-spacing) as automatic rules; and (iii)~a dependency-free
implementation in which documents remain plain data (\texttt{.jgb.json}),
enabling print-quality output, playback, and interchange with recently
proposed machine-oriented encodings of \term{jeongganbo}. We specify the text
format and its rhythmic semantics, describe the engraving model and its
treatment of typographic edge cases (Unicode gaps in octave-variant
Sinographs, ink-ratio compensation between glyphs and vector symbols), and
discuss limitations and planned interoperability with MusicXML/MEI and optical
music recognition pipelines. The implementation was produced through an
AI-assisted workflow led by a domain expert and verified behaviorally
against published reference scores; this workflow and its verification
regime are disclosed in the report.

\end{abstract}

\section{Introduction}\label{sec:intro}

\term{Jeongganbo} is the standard notation of Korean traditional (court and
educational) music. A score is a lattice of square cells (\term{jeonggan},
井間, ``well-space'' --- the grid lines trace the character 井, `well');
each cell is one beat, a vertical run of cells forms a column
(\term{gak}, 刻) that typically carries one rhythmic cycle, and columns
proceed from right to left. Pitch is written inside cells using the twelve
pitch names (\term{yulmyeong}, 律名), and duration is not written with
symbols at all. The notation's central move is to spatialize time: each beat
becomes an area --- the cell --- and a pitch's duration is expressed by
\emph{where it sits and how much of that area it claims}. A cell
holding one syllable sounds for the whole beat; a cell split into two rows
halves it; two syllables side by side within a row subdivide that row's span.
The cell is at once a ruler and a container: alongside pitch, it anchors the
rhythm-cycle, playing-technique, and lyric elements that share its frame.
\term{Jeongganbo} is regarded as the earliest mensural notation in East
Asia~\cite{ngc2010notations} --- duration was written into the score as
early as the fifteenth century.
Its structure, however, differs at the most basic level from the staff model
that mainstream engraving software assumes, and realizing a grid,
right-to-left vertical flow, duration by position and share,
\term{yulmyeong} pitch names, and the ornament vocabulary
(\term{sigimsae}, \kr{시김새}) of published scores on top of such tools is
impractical.

The practical consequence is well documented: scores are typeset by hand in
word-processor tables, produced as images rather than data, and rebuilt from
scratch each time. A series of dedicated tools appeared between 2001 and
2024, but none combined input, full ornament typesetting, traditional
page conventions, print output, and playback, and the installable ones are no
longer distributed (\S\ref{sec:related}).

This report documents the technical design of \term{Umulsai}, a
\term{jeongganbo} editor deployed as a static web page. The name renders the
two Sinographs of \term{jeonggan} --- 井 ``well'' and 間 ``between'' --- into
native Korean (\term{umul} + \term{sai}). Contributions:

\begin{enumerate}
\item \textbf{A plain-text notation format} for \term{jeongganbo} in which the
text's spatial arrangement maps one-to-one onto the notational grid,
preserving the notation's own position-as-duration principle in an editable,
diffable, human-writable string (\S\ref{sec:format}). The format is
structurally convergent with the machine-oriented encodings that recent MIR
work on \term{jeongganbo} arrived at independently, which makes round-tripping
with recognition and generation pipelines a mapping-table exercise rather than
a redesign (\S\ref{sec:interop}).
\item \textbf{An engraving model} that turns the conventions of published
scores --- right-to-left band layout, \term{daegang} dividing lines, the
traditional title column, the vertical tempo formula, ornament attachment, and
several micro-typographic habits of professional engravers --- into explicit,
automatic rules (\S\ref{sec:engraving}).
\item \textbf{A serverless implementation}: one static page, no installation,
no account, no server round-trips; documents are autosaved locally and
exchanged as a single JSON file (\S\ref{sec:impl}).
\end{enumerate}

The implementation was AI-written under natural-language specification by a
domain expert; \S\ref{sec:vibecoding} discloses this workflow and the
verification regime it requires.

The system has been publicly deployed at \url{https://www.umulsai.com} since
July 2026. An
accompanying study of user needs and a preliminary usability evaluation are
reported separately (in preparation); the present report is
confined to the technical design.

\section{Structural Requirements of \term{Jeongganbo}}\label{sec:background}

We summarize the notational elements a digital tool must carry, as they define
the design space.

\paragraph{Grid and time.} The beat count $n$ of the accompanying rhythmic
cycle (\term{jangdan}, \kr{장단}) carried by one \term{gak} is a per-document
constant (4, 6, 8, 10, 12, 16, and 20 are common). Cells are grouped by thicker horizontal lines into
sections (\term{daegang}, 大綱), e.g.\ $3{+}3{+}3{+}3$ for 12 beats or
$11{+}5$ for 16. Reading order is: within a cell, rows top-to-bottom and, within a row,
left-to-right; within a column, top-to-bottom; across columns, right-to-left.

\paragraph{Pitch.} Twelve pitch names (黃大太夾姑仲\rgYubin{}林夷南無應), absolute,
with \term{hwangjong} (黃鍾) conventionally near E$\flat$4 in modern practice.
Octave displacement is written not with numerals but by modifying the
Sinograph itself: one octave up adds the water radical (\rgSamsu{}) --- 黃→潢; two up
doubles it (黃→\rgHwangUpTwo{}); one down adds the person radical (\rgInbyeon{}) --- 黃→\rgHwangDownOne{}; two
down uses the step radical (\rgDuin{}) --- 黃→\rgHwangDownTwo{}. In Hangul the prefixes
\term{cheong}/\term{jungcheong} (up) and \term{bae}/\term{habae} (down) serve
the same role. A digital tool must treat these as systematic octave marks, not
as arbitrary characters --- including the two combinations whose variant
Sinographs do not exist in Unicode (\S\ref{sec:conventions}).

\paragraph{Ornaments.} Published scores attach a rich vocabulary of
\term{sigimsae} signs: small marks placed to the right of a note (vibrato,
slides, appoggiatura families), full-cell signs, and fingering or technique
marks. These are graphical symbols with conventional shapes, standardized in
practice by the National Gugak Center's instrumental score
series~\cite{ngcjeongak}; any
faithful engraving must reproduce them and their placement.

\paragraph{Parallel layers.} Alongside the melody column run (a)~a
\term{jangdan} column of drum mnemonics (\term{deong}, \term{gideok},
\term{kung}, \dots) written once at the head of the piece, and (b)~a narrow
annotation lane at the right edge of each melody cell carrying lyrics,
bowing/technique signs, or drum mnemonics. Layers are aligned cell-by-cell;
structural edits (inserting or deleting a \term{gak}) must shift all layers
together.

\paragraph{Page conventions.} Published scores open with a vertical title
column at the right edge of the first page, mark tempo with a vertical
Sinograph formula (一分·六十井 --- ``60 cells per minute'', possibly a
range), label sections above columns (章 numbering, 大餘音, \dots), and are
printed on fixed paper sizes with the grid sized in physical units.

These yield the requirements: \textbf{R1} positional duration model;
\textbf{R2} systematic octave-variant rendering; \textbf{R3} full ornament
typesetting; \textbf{R4} multi-layer cell alignment; \textbf{R5} automation of
page conventions in physical units.

\section{Related Work}\label{sec:related}

\paragraph{Encodings.} An XML document-type definition for \term{jeongganbo}
symbols was proposed in 2013 and extended in a 2017 patent
\cite{lee2013xml,lee2017patent}. Recent MIR work encodes \term{jeongganbo} as
plain text: the \emph{Six Dragons} project encodes fifteenth-century court
sources for masked-language-model reconstruction \cite{han2026sixdragons}, and
a \term{jeongganbo} OMR system outputs tokens where a line is a \term{gak}, a
delimiter separates cells, and an index encodes position within the cell
\cite{kim2025omr}. These encodings, however, are machine-oriented:
they are produced and consumed by models, and neither is intended to be
written by hand or engraved back to a page --- their structural convergence
with the present format is taken up in \S\ref{sec:interop}.

\paragraph{Editors.} Input systems of 2001 stored pitch glyphs at free
coordinates or targeted a single instrument with MIDI conversion
\cite{park2001input,kim2001system}. A 2010 desktop authoring tool reached
lyric/percussion co-notation but left \term{daegang} lines to be drawn
manually and had no playback \cite{lee2010authoring}; it and its
contemporaries are no longer distributed. A 2022 design study proposed a
server- and membership-based web editor with a fixed six-way cell split
\cite{shim2022design}. An open web editor (2024) provides grid input but no
ornaments, lyrics, or print layout \cite{jgbeditor}. No prior system covers
R1--R5; ornament typesetting (R3) in particular has been absent or vestigial
throughout.

\paragraph{Non-Western notation systems generally.} The problem class ---
reviving a non-CWMN notation as data plus engraving --- is active
internationally: MEI workflows for Ottoman Hamparsum sources \cite{olley2019}
and Japanese court music \cite{seki2023}, OMR datasets for Song-dynasty
\term{suzipu} \cite{repolusk2024}, and the SymbTr corpus for Turkish makam
music, which grew from a bespoke text format into a multi-thousand-piece
research resource \cite{karaosmanoglu2012}. SymbTr in particular illustrates
the trajectory this system aims at: a human-usable format that later serves as
corpus infrastructure.

\section{The Text Format}\label{sec:format}

\subsection{Syntax}\label{sec:syntax}

\begin{figure}[htbp]
\centering
\includegraphics[width=.9\linewidth]{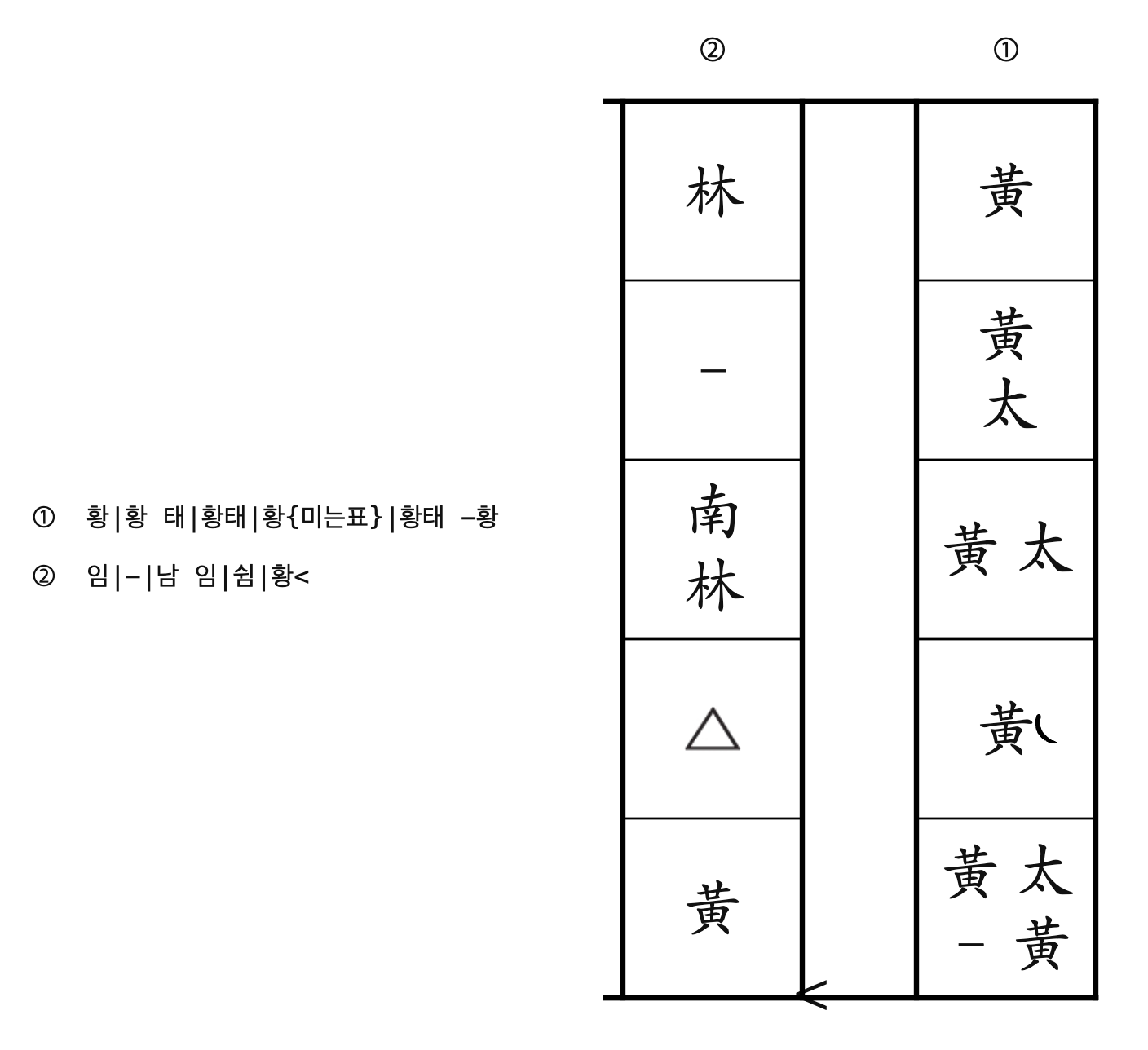}
\caption{Text--grid correspondence. Two input lines (left) and the engraved
result (right; columns read right-to-left, so line \textcircled{1} is the
rightmost column). Line \textcircled{1} exercises single note, row division,
intra-row adjacency, ornament token, and sustain; line \textcircled{2} adds
rest and breath mark.}
\label{fig:mapping}
\end{figure}

A document's melody is a single string. Its structure is given in
Table~\ref{tab:syntax}; Figure~\ref{fig:mapping} shows the correspondence
between a two-\term{gak} string and its engraved result.

\begin{table}[htbp]
\centering\small
\caption{Melody string constructs.}
\label{tab:syntax}
\begin{tabular}{ll}
\toprule
Construct & Meaning \\
\midrule
line break & next \term{gak} (column) \\
\texttt{|} & next \term{jeonggan} (cell) within the column \\
whitespace inside a cell & row division --- the beat is split vertically \\
adjacency (no space) & placement side-by-side within the same row \\
\kr{황 대 태 \dots\ 응} & the twelve \term{yulmyeong}, written in Hangul \\
\kr{청·중청} / \kr{배·하배} prefix & octave $+1/{+}2$ and $-1/{-}2$ \\
\texttt{-} & sustain of the previous note \\
\kr{쉼} & rest \\
\texttt{<} suffixed to a note & breath mark (rendered at the cell's lower right) \\
\texttt{\{}name\texttt{\}} after a note & ornament attachment (also \texttt{[}name\texttt{]}, \texttt{(}name\texttt{)}) \\
\texttt{\{}name\texttt{@}s\texttt{,}dx\texttt{,}dy\texttt{\}} & ornament with size (\%) and offset (\% of cell) adjustments \\
\bottomrule
\end{tabular}
\end{table}

Example: \texttt{\kr{황 태협|협<|임}} denotes three cells; the first has two
rows --- 黃 alone, then 太·夾 side by side --- the second 夾 with a breath
mark, the third 林.

The three bracket styles are equivalent on input and normalized to
\texttt{\{\}} on serialization; this tolerance exists because keyboard layouts
differ in which bracket is cheapest to type. Ornament names are the Korean
display names themselves (e.g.\ \texttt{\{\kr{미는표}\}}); symbols whose
official names are unsettled carry provisional identifiers
(\texttt{\{s11\}}), so a later renaming changes only the display table, not
documents' semantics. Tokens that have already been renamed remain readable
through a read-only alias table (\texttt{LEGACY\_ALIAS}), so existing
documents open unchanged; the alias table is a closed list that accepts no
new entries, to keep the compatibility layer from blurring the name space.

\begin{figure}[htbp]
\centering
\includegraphics[width=.7\linewidth]{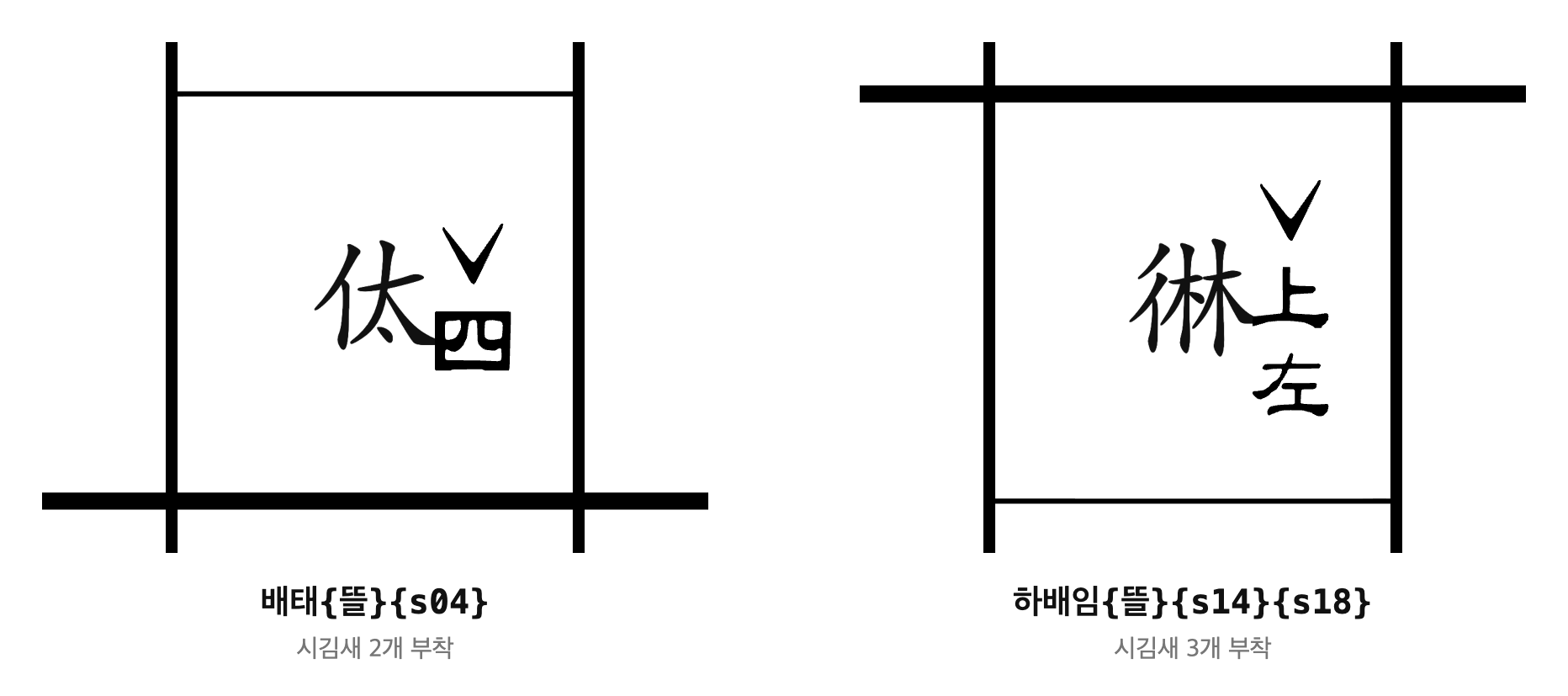}
\caption{Multiple ornaments attached to one note (actual engraved output).
Left: two ornaments (\texttt{\kr{배태}\{\kr{뜰}\}\{s04\}}). Right: three
ornaments (\texttt{\kr{하배임}\{\kr{뜰}\}\{s14\}\{s18\}}) --- chained tokens
stack in order beside the note.}
\label{fig:ornstack}
\end{figure}

Ornament tokens may be chained after a single note
(Figure~\ref{fig:ornstack}): the published-score practice of stacking two or
three marks on one pitch engraves directly within a single cell, and combines
naturally with the low-octave prefix glyphs (\kr{배}·\kr{하배}).

\subsection{Rhythmic semantics}\label{sec:semantics}

Let a cell contain $r$ whitespace-separated rows. Each row receives $1/r$ of
the beat. Within a row, the note-groups (a pitch plus its suffixed marks and
ornament tokens form one group) divide the row's span equally. A cell that is
empty, or a group consisting of \texttt{-}, extends the previous sounding
note; \kr{쉼} is silence. Breath marks and tempo-class ornament tokens occupy
no time slot. This is the complete duration model: there are no duration
symbols anywhere in the format, satisfying R1 by construction.

Two properties follow. First, \textbf{every prefix of a valid document is a
valid document}: the grid is always drawn complete from the document's meter
parameter (cells per \term{gak}), so unfinished input still renders as a
well-formed score --- there is no ``syntax error'' state during entry, only
emptier cells. Second, the format is \textbf{local}: editing one cell's text
cannot change the interpretation of any other cell, which makes the format
diff-friendly and makes incremental re-engraving trivial.

\subsection{Pitch model}\label{sec:pitch}

The twelve \term{yulmyeong} map to semitone indices with \term{hwangjong}
$=0$; octave prefixes add $\pm12/\pm24$. Playback tuning is equal-tempered
from a configurable \term{hwangjong} reference (E$\flat$4 by default, C4
selectable) --- a deliberate simplification; just intonation and
instrument-specific temperament are out of scope for the current build
(\S\ref{sec:limits}). Display is switchable between Sinograph and Hangul ---
octave-variant glyph selection, and the image-based typesetting of extended
low-register names with no Unicode representation (e.g.\ \kr{하하배임}), are
taken up in \S\ref{sec:conventions}.

\subsection{Ornament tokens as the single source of truth}\label{sec:orntoken}

\begin{figure}[htbp]
\centering
\includegraphics[width=.85\linewidth]{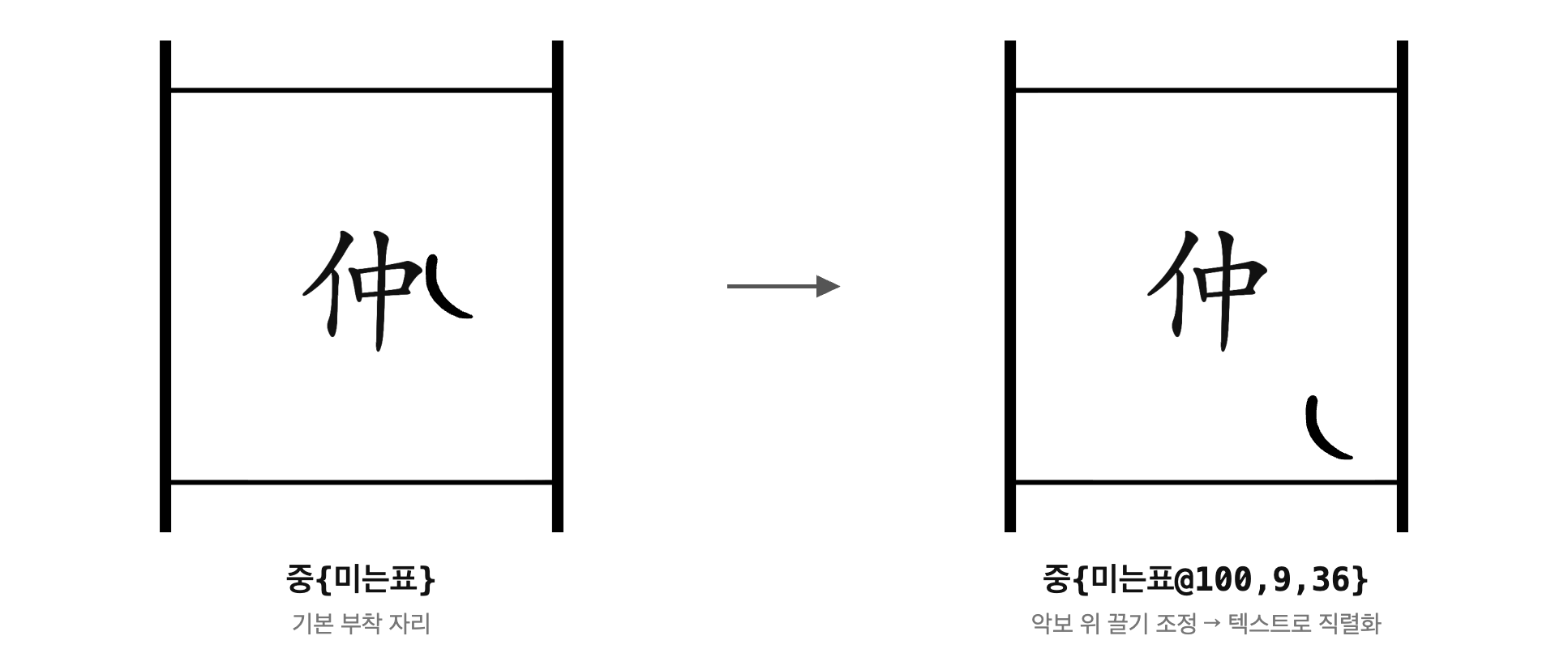}
\caption{Engraving adjustments serialized into the notation text. Left: the
ornament \kr{미는표} at its default attachment on 仲. Right: after dragging
on the score, the adjustment is stored in the token itself
(\texttt{@100,9,36} = size unchanged at 100\,\%, offset $+9/{+}36$\,\% of the
cell), so the page remains a pure function of the document string.}
\label{fig:orn}
\end{figure}

Ornament placement in published scores is a typographic judgment: engravers
nudge and scale marks to avoid collisions. \term{Umulsai} lets the user drag
and resize any adjustable ornament directly on the score; the adjustment is
then serialized \emph{into the melody text} as
\texttt{\{}name\texttt{@}size\texttt{,}dx\texttt{,}dy\texttt{\}} (percent
units relative to the cell). The rendered page is therefore a pure function of
the document string plus layout parameters --- there is no hidden layout
state, and a document pasted into another instance reproduces its page
exactly. Default-valued adjustments serialize back to the bare
\texttt{\{}name\texttt{\}} (Figure~\ref{fig:orn}).

\subsection{Convergence and interoperability}\label{sec:interop}

The format is structurally aligned with the OMR output tokens of
Kim et al.\ \cite{kim2025omr} (line = \term{gak}, \texttt{|} = cell,
intra-cell position index) and with the \emph{Six Dragons} encoding of
Han et al.\ \cite{han2026sixdragons}. The differences are surface-level
(position indices vs.\ literal whitespace layout; symbol vocabularies). A
deterministic mapping table suffices to import OMR output for post-correction
and engraving, or to export corpus data for model training --- the workflow
gap both papers leave open (recognition accuracy short of publication quality;
generation output needing human editing) is exactly an editor. The mapping
table itself is not implemented as of this report; it is planned with the
OMR output tokens of Kim et al.\ as the primary target.

\section{The Engraving Model}\label{sec:engraving}

\subsection{Page model}\label{sec:page}

\begin{figure}[htbp]
\centering
\includegraphics[width=.8\linewidth]{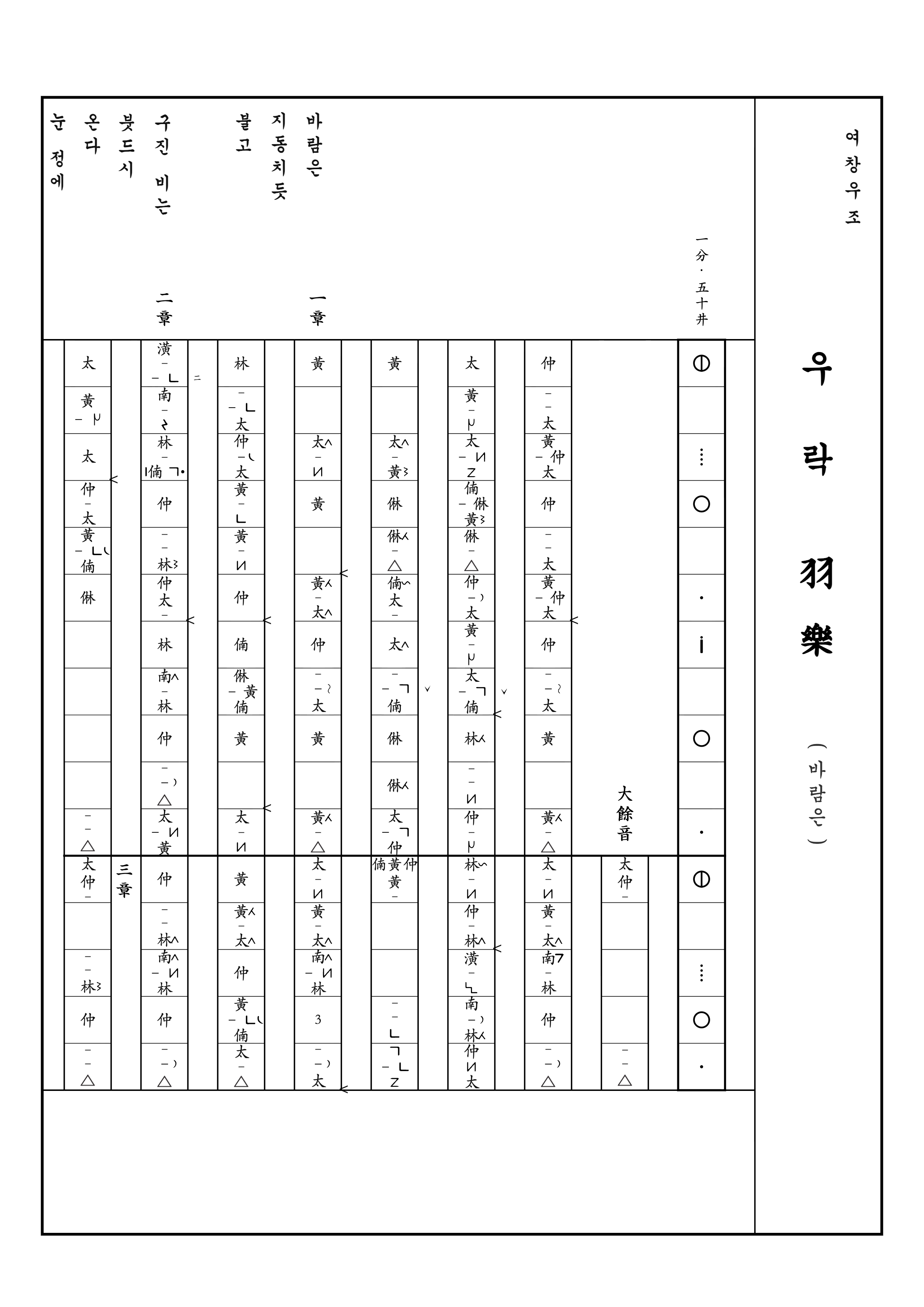}
\caption{The opening page of \emph{Urak} (\kr{우락} 羽樂, female-voice
\term{ujo} \term{gagok}) engraved by \term{Umulsai}: a 16-cell grid,
traditional vertical title column at the right edge, \term{jangdan} column
with drum mnemonics, vertical Sino-Korean tempo mark (一分·五十井), section
names above the columns (\kr{대여음}, sections 1--3), lyric annotation lane,
per-note
ornaments, right-to-left band flow.}
\label{fig:page}
\end{figure}

Engraving targets physical units end-to-end: the page is A4 (portrait or
landscape), the grid is computed in millimetres, and screen rendering, print,
and PNG export share one SVG generator per page. Layout parameters: cells per
\term{gak} (1--64), \term{gak} per band (1--40), bands per page (1--12 or
automatic), cell width (4--30\,mm), inter-column and inter-band gaps, global
scale. The traditional title column occupies a whole number of \term{gak}
slots (1--4 or automatic) on the first page, so first-page capacity is
$(\textit{gak per band} - \textit{title slots}) \times \textit{bands} -
\textit{jangdan slot}$; keeping the title width in \term{gak} units rather
than millimetres preserves grid alignment by construction. If the configured
grid exceeds the printable area, the whole layout is shrunk proportionally and
the effective scale is reported to the user. Figure~\ref{fig:page} shows a
complete engraved page of a real piece --- a daegeum transcription of the
\term{gagok} \emph{Urak}.

\subsection{Formalized engraving conventions}\label{sec:conventions}

\begin{figure}[htbp]
\centering
\includegraphics[width=\linewidth]{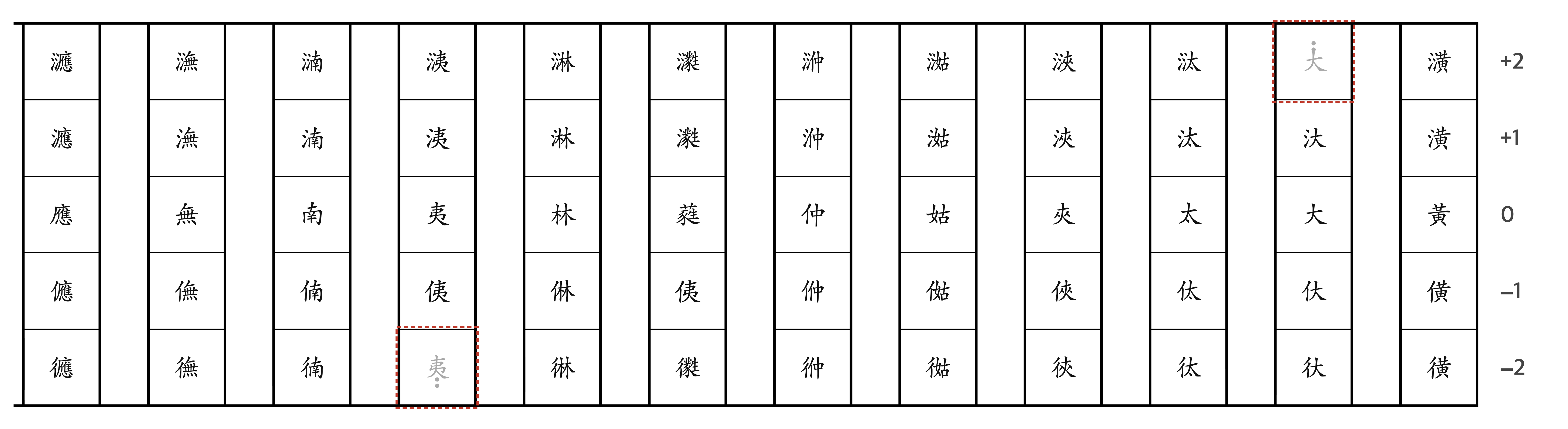}
\caption{The full octave-variant inventory as engraved by the system: twelve
\term{yulmyeong} (columns, right-to-left from 黃) across five octaves (rows,
$+2$ to $-2$). Dashed boxes mark the two combinations with no Unicode
codepoint ($+2$ of 大, $-2$ of 夷), which fall back to the base glyph with
conventional octave dots in subdued gray.}
\label{fig:octave}
\end{figure}

The interesting part of the engraving model is the codification of practices
that human engravers apply silently. The rules below are defaults, not
options:

\begin{itemize}
\item \textbf{Right-to-left flow.} Bands fill right to left; page order,
column numbering, and the title column position all follow.
\item \textbf{\term{Daegang} lines.} Thick horizontal rules at the section
boundaries, with presets per meter ($10 = 3{\cdot}2{\cdot}2{\cdot}3$,
$12 = 3{\cdot}3{\cdot}3{\cdot}3$, $16 = 11{\cdot}5$,
$20 = 6{\cdot}4{\cdot}4{\cdot}6$) and arbitrary user partitions whose sum must
equal the meter.
\item \textbf{Sustain-row compression.} In published scores a subdivision row
containing only the sustain stroke is visually de-emphasized: the row's
vertical share is reduced ($\times0.68$), the stroke drawn slightly smaller
($\times0.85$) but stretched horizontally ($\times1.95$) so it still reads as
a continuation, and remaining space is redistributed to the sounding rows,
which drift toward the cell's center. \term{Umulsai} applies this
automatically.
\item \textbf{Two-note tightening.} A cell containing exactly two notes
(either two rows or one row of two) has its internal spacing tightened by
20\,\% to avoid a straggling appearance.
\item \textbf{Ink-ratio compensation.} CJK glyphs fill roughly 0.86 of their
nominal em square with ink (measured on 林, 黃), whereas ink-cropped vector
symbols fill ${\sim}0.96$ of their box; naively equal sizes make images look
larger than neighboring glyphs. Image-based note names are scaled by 0.9 to
equalize perceived size.
\item \textbf{Octave-variant glyph selection.} Sinograph display uses the
radical-modified variant characters across $\pm2$ octaves
(Figure~\ref{fig:octave}). Two combinations (+2 of 大, $-2$ of 夷) have no
Unicode codepoint; these fall back to the base glyph plus a conventional
octave dot, rendered in a subdued gray so the substitution is visible but
unobtrusive.
\item \textbf{Vertical text.} Titles, section labels, and the tempo formula
are set vertically; enclosing brackets rotate 90° clockwise per traditional
vertical composition, while each rotated glyph still occupies one character
slot so the vertical rhythm of the column is preserved.
\item \textbf{Tempo formula.} Tempo is written as 一分·N井 (``N cells per
minute''), supporting ranges with the full-width tilde (一分·六十$\sim$八十井)
--- the half-width tilde reads as a stray stroke in vertical setting --- and
following the score-house convention of omitting 百 in non-round hundreds
($160 \rightarrow$ 一六十) while keeping it for round ones (一百).
\end{itemize}

\subsection{Ornament model}\label{sec:ornmodel}

The symbol inventory is a declarative registry: each entry carries an
identifier, display name, and a map from \emph{placement sites} to default
scales. Sites are: attached (small, right of a note), cell (occupying a whole
\term{jeonggan}), tempo (vertical, outside the cell's right edge),
annotation-lane, and \term{jangdan}-lane. The current build registers
\textbf{84 distinct \term{sigimsae} signs} --- 64 attached marks (including
fingering marks), 18 full-cell signs, and 2 usable in both positions --- plus
5 tempo indications, for 89 melody-palette symbols in total, and drum-mnemonic
glyphs usable in both the \term{jangdan} column and the annotation lane. These
counts are deduplicated: no glyph or name appears twice within the palette
(across all lanes the registry holds 103 site-entries over 102 unique graphics
--- the single shared graphic, the hold sign, is registered once per lane and
counted once here). Per-site scales are set visually, not proportionally: the
same symbol may need different absolute sizes in lanes of different widths to
read as the same-sized mark.

Because placement sites, names, and scales live in one registry, the three
symbol palettes (melody, annotation lane, \term{jangdan}) are generated by one
builder, and a symbol admitted to a new site inherits search, recency, and
tooltip behavior without new code.

\subsection{Layer alignment and structural edits}\label{sec:layers}

Melody, annotation lane, and \term{jangdan} are aligned per \term{gak} and per
cell. Structural operations (insert/delete \term{gak}, clipboard operations
over a selected cell range) shift all layers together; the \term{jangdan}
column, being a single pattern written once, is excluded from range operations
by design. Cell styling (background fills for pedagogical highlighting, merged
cells, erased cells, per-edge rule styles) is stored per cell and follows the
same shifts. Erasing and merging are implemented as white masks over the
ruling, but boundary lines shared with neighboring cells and the structural
skeleton (outer frame, \term{daegang} lines) are re-drawn on top of any mask,
so emptying one cell never opens its neighbor's wall or breaks a
\term{daegang} line.

\section{Implementation}\label{sec:impl}

\begin{figure}[htbp]
\centering
\includegraphics[width=\linewidth]{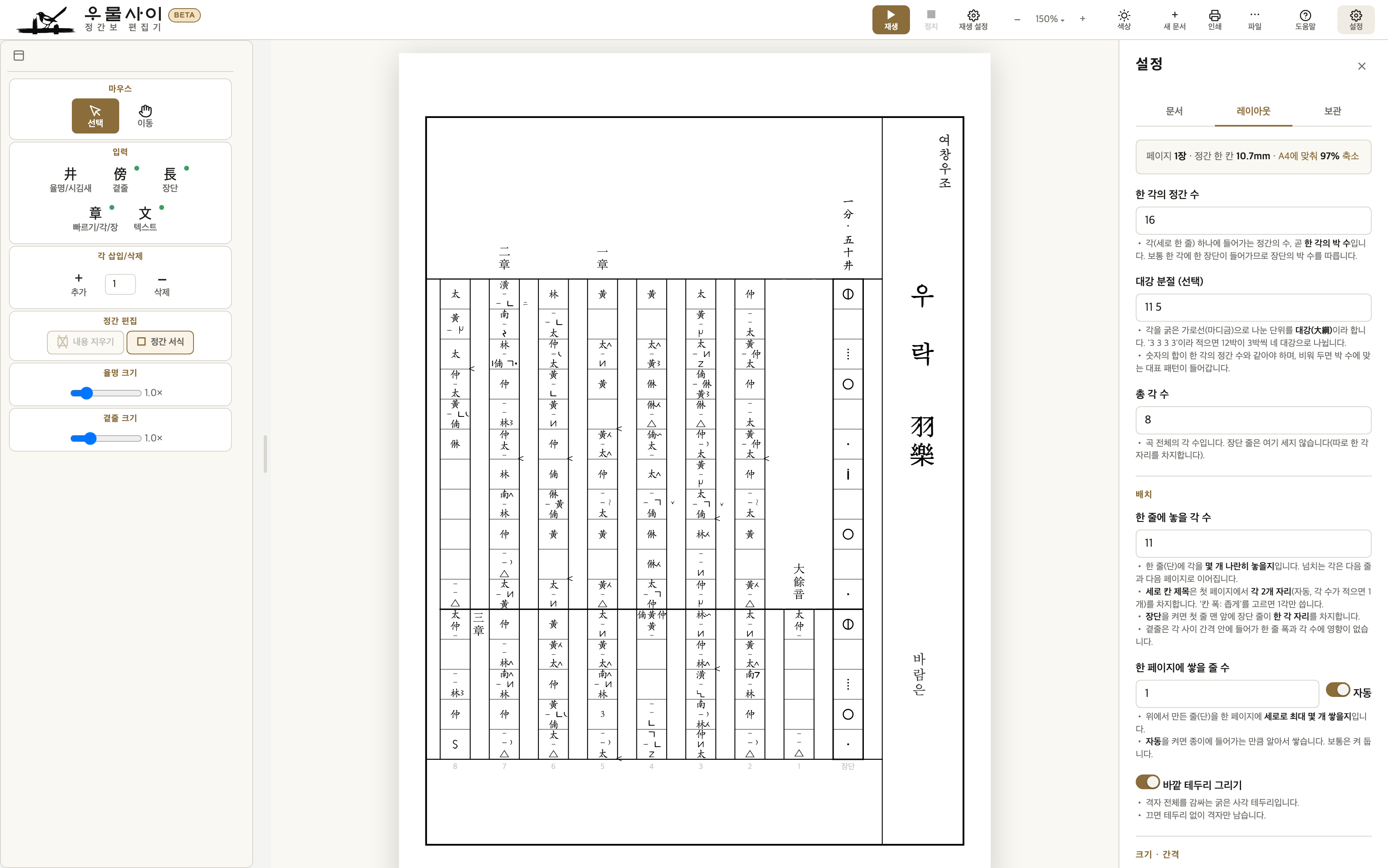}
\caption{The \term{Umulsai} editing screen (the \emph{Urak} document,
direct-input mode): the docked toolbar on the left (input tools, column
insertion/deletion, cell formatting), the A4 page preview in the center, and
the settings sidebar on the right. The layout tab reports the physical
result --- cell size 10.7\,mm, shrunk to 97\,\% to fit A4
(\S\ref{sec:page}).}
\label{fig:ui}
\end{figure}

\term{Umulsai} is a static web application: one HTML page, one stylesheet, and
vanilla JavaScript (${\sim}6{,}800$ lines of application code plus a
${\sim}260$-line symbol registry) with \textbf{no runtime dependencies, no
build step, no server, and no account system}. All notation graphics --- the
\term{yulmyeong} image set, ornament SVGs, and drum glyphs, ${\sim}500$\,KB
in total ---
are inlined as data URIs, so the deployed artifact is self-contained and works
from a local file. Documents autosave to \texttt{localStorage} on every edit;
the interchange format \texttt{.jgb.json} serializes the full document
(settings, melody string, layer strings, cell styles, text objects). Output
paths: browser print (PDF) with print CSS that strips all UI, and per-page PNG
rasterized at 300\,dpi. Playback synthesizes sine tones from the event list
derived by the semantics of \S\ref{sec:semantics} (20--240\,BPM), highlighting
the sounding cell; the sounding BPM and the notated tempo formula are
deliberately decoupled once the user overrides the former, since a notated
range has no single playable value. Figure~\ref{fig:ui} shows the editing
screen.

The serverless constraint is a distribution argument as much as an engineering
one: every prior installable \term{jeongganbo} tool died with its distribution
channel, and a static page with no backend has no server to decommission. It
also eliminates accounts and telemetry as privacy surface --- the analytics
layer is a no-op stub unless explicitly wired to a cookie-less counter.

\subsection{Development methodology: AI-assisted implementation,
expert-verified behavior}\label{sec:vibecoding}

\paragraph{Disclosure.} The implementation was produced predominantly through
an AI-assisted workflow --- the practice now commonly called \emph{vibe
coding}: the author, a domain practitioner rather than a professional software
engineer, specified behavior, notational rules, and corrections in natural
language, and LLM coding agents wrote and revised the code. The author does
not claim line-level authorship of the codebase and takes responsibility for
the system at the level of its specified and verified behavior. This statement
follows the disclosure norms that publishers and venues converged on in
2025--2026, under which generative-AI contributions to code are reported
separately from writing assistance, with the human author retaining full
responsibility for the result.

\paragraph{Epistemic stance of this report.} Because the code was AI-written,
the claims in this report are deliberately confined to properties that are
\emph{externally verifiable from the artifact}: the grammar and semantics of
the text format (\S\ref{sec:format} --- checkable by typing), the engraving
rules (\S\ref{sec:engraving} --- checkable by comparing rendered pages against
published scores), the file format, and measurable implementation facts (size,
dependency count, offline operation). No claim is made about internal code
quality or architectural elegance.

\paragraph{Verification in place of line-level review.} The known risk profile
of vibe-coded software is reduced code inspection and skipped quality
assurance. Here, assurance was shifted to the behavioral level, where the
author \emph{is} the qualified reviewer:

\begin{enumerate}
\item \textbf{Reference-score acceptance testing.} Every engraving rule of
\S\ref{sec:conventions} was iterated against reference pages from the National
Gugak Center \term{jeongak} series until the rendered output was judged
faithful at print size; the micro-typographic constants (sustain-row 0.68,
ink-ratio 0.9, \dots) are fitted to those references, not chosen a priori.
\item \textbf{A regression harness for notational integrity.} A standalone
script re-derives the nine symbol-placement tables from the declarative
registry and compares them against a known-good baseline commit; additions
pass, but any silent change to an existing symbol's name, site, or scale ---
the kind of drift AI edits can introduce --- fails the check.
\item \textbf{Specification-of-record documentation.} Design rationale (why
each rule exists, what must not be regressed) is co-maintained with the code
as a standing document that every AI session reads before editing. In this
workflow the durable artifact of authorship is the specification and its
rationale; code is its replaceable projection.
\item \textbf{Architecture as risk containment.} The dominant security
concerns of AI-generated web code --- authentication, server-side injection,
data exfiltration --- are absent by construction, as \S\ref{sec:impl}
described: there is no network I/O beyond fetching the page itself.
\end{enumerate}

\section{Comparison}\label{sec:comparison}

As \S\ref{sec:related} showed, no prior system integrates R1--R5, and the de
facto standard --- word-processor tables --- produces images rather than
data, leaving every convention of
\S\ref{sec:conventions} as manual labor. To our knowledge, \term{Umulsai}
is the first \term{jeongganbo} system to combine a human-writable text format,
full ornament typesetting, automated traditional page conventions,
physical-unit print output, and playback in one artifact; it is also, we
believe, the first to serialize engraving adjustments into the notation text
itself (\S\ref{sec:orntoken}). The main capability found in prior work but
\emph{not} here is MIDI file export \cite{shim2022design}.

\section{Limitations and Future Work}\label{sec:limits}

\begin{itemize}
\item \textbf{Playback fidelity.} Sine-wave monophony; \term{sigimsae} (which
are precisely pitch inflections) and \term{jangdan} strokes are not sounded.
Assigning acoustic rules to the ornament vocabulary is an independent research
task; the serverless constraint favors lightweight synthesis over sample
libraries.
\item \textbf{Temperament.} Equal-tempered playback from a movable
\term{hwangjong}; historical/instrumental intonation is unaddressed (a
limitation already noted in the 2001 MIDI work \cite{kim2001system}).
\item \textbf{Interchange.} No MusicXML/MEI/MIDI export yet. The OMR/corpus
mapping of \S\ref{sec:interop} is specified but not implemented.
\item \textbf{Coverage.} The symbol inventory follows the National Gugak
Center \term{jeongak} series; folk-music (\term{minsogak}) notation practices
and instrument-specific extensions (e.g.\ dual fret notation for
\term{geomungo}) are future inventory work.
\item \textbf{Assurance.} Beyond the registry regression check
(\S\ref{sec:vibecoding}), there is no automated test suite; correctness
assurance is behavioral --- domain-expert acceptance against reference scores
--- rather than code-review- or proof-based. This is an honest consequence of
the AI-assisted development model and is partially contained by the serverless
architecture; systematic tests over the format's semantics
(\S\ref{sec:semantics}) are a natural next step.
\item \textbf{Evaluation.} This report is a system description; user-facing
evaluation is reported separately (in preparation).
\end{itemize}

\section{Availability}\label{sec:availability}

The system is deployed at \url{https://www.umulsai.com} (no installation or
account; Korean UI). This report describes the July 2026 build. The source
code is available on GitHub under the MIT license
(\url{https://github.com/danbinaerinHan/jeongganbo-generator}), and the build
this report describes is archived as a Zenodo snapshot
(DOI: \url{https://doi.org/10.5281/zenodo.21616981}).

\end{document}